# IEEE Copyright Notice

# Ultra-broadband integrated optical filters based on adiabatic optimization of coupled waveguides

Kazim Gorgulu, Emir Salih Magden

*Abstract*—Broadband spectral filters are highly sought-after in many integrated photonics applications such as ultra-broadband wavelength division multiplexing, multi-band spectroscopy, and broadband sensing. In this study, we present the design, simulation, and experimental demonstration of compact and ultra-broadband silicon photonic filters with adiabatic waveguides. We first develop an optimization algorithm for coupled adiabatic waveguide structures, and use it to design individual, single-cutoff spectral filters. These single-cutoff filters are 1×2 port devices that optimally separate a broadband signal into short-pass and long-pass outputs, within a specified device length. We control the power roll-off and extinction ratio in these filters using the adiabaticity parameter. Both outputs of the filters operate in transmission, making it possible to cascade multiple filters in different configurations. Taking advantage of this flexibility, we cascade two filters with different cutoff wavelengths on-chip, and experimentally demonstrate band-pass operation. The independent and flexible design of these band edges enables filters with bandwidths well over 100 nm. Experimentally, we demonstrate band-pass filters with passbands ranging from 6.4 nm up to 96.6 nm. Our devices achieve flat-band transmission in all three of the short-pass, band-pass, and long-pass outputs with less than 1.5 dB insertion loss and extinction ratios of over 15 dB. These ultra-broadband filters can enable new capabilities for multi-band integrated photonics in communications, spectroscopy, and sensing applications.

*Index Terms*—adiabatic optimization, ultra-broadband filters, photonic integrated circuits, silicon photonics

## I. Introduction

INTEGRATED optical filters are one of the main building blocks in photonic integrated circuits and are commonly used in communications [1-4], spectroscopy [5-7], and sensing [8-10]. Many of these applications require high-performance and wideband optical filters for the development of next-generation broadband photonic systems. Efficiently splitting and combining ultra-broadband signals is crucial in wideband wavelength division multiplexing to enable transmission beyond the typical communication bands to increase optical networks' transport capacities [3, 4, 11, 12]. Broadband filtering can also be utilized for increasing the operation bandwidth of narrowband spectroscopy [5, 13, 14], or separating different spectral components in dual-comb spectroscopy [15] to enhance measurement resolution and signal-to-noise ratio.

So far, many different structures have been presented for implementing integrated optical filters. Among these, ring resonators [16-18], Mach-Zehnder interferometers (MZIs) [19-21], arrayed waveguide gratings (AWGs) [22, 23], and echelle gratings [24, 25] have been studied extensively. As these structures achieve spectral filtering through interference and waveguide dispersion, their bandwidths can typically reach about 30 nm [25, 26], limiting their usability in applications requiring wider passbands. Due to their limited operation bandwidths, these filters are primarily used in narrowband spectral filtering and multiplexing applications [26]. Cascaded ring resonators [27] and subwavelength grating (SWG)-assisted MZIs [28] have also been demonstrated with flat-top and broader-band operation but have only achieved up to about 40 nm-wide bandwidths so far. In general, system design with these interferometric filters requires careful consideration of their free spectral ranges (FSR), especially for broadband filtering [26]. On the other hand, Bragg grating-based optical filters can be designed with FSR-free operation [29, 30]. However, these Bragg grating filters work in reflection mode and thus require an additional optical circulator or directional coupler for signal routing, significantly increasing insertion loss and integration complexity. In comparison, Bragg grating-based contra-directional couplers can allow for add-drop filtering configurations without requiring optical circulators [31-33]. Yet, despite their good extinction ratios, these filters have so far only achieved optical bandwidths up to 35 nm, due to the limited coupling strength of forward and backward propagating modes. More recently, a broader-band design of the contra-directional Bragg grating filter was demonstrated by chirping the grating period [34], with passbands reaching up to 11 THz (approximately 88 nm). However, this method results in much longer device lengths, and presents an inherent design tradeoff between extinction ratio and device bandwidth.

Unlike these grating-based or typical interferometric devices, couplers based on adiabatic waveguide transitions present an important class of integrated photonic devices that are particularly suitable for broadband operation. An adiabatic coupler is typically designed to maintain the propagating optical signal in a single eigenmode through a waveguide

This work is supported by Scientific and Technological Research Council of Turkey (TUBITAK) under grant number 118E159. (Corresponding author: Emir Salih Magden).

The authors are with the Department of Electrical and Electronics Engineering, Koç University, Sarıyer, İstanbul 34450, Turkey. (e-mail: kgorgulu18@ku.edu.tr, esmagden@ku.edu.tr)

whose cross-section slowly evolves in the propagation direction [35-37]. While many adiabatic devices have been demonstrated for broadband polarization rotation [38, 39], mode multiplexing [40, 41], and power splitting [42, 43], adiabatic couplers can also be used for on-chip broadband filtering [44-47]. These adiabatic filters rely on the phase mismatch between coupled waveguide modes at different wavelengths while achieving broadband operation through slowly varying waveguide cross-sections. However, these filters must be made sufficiently long to satisfy the desirable extinction ratio and roll-off performance. Typical lengths of these filters vary from several millimeters to centimeters, significantly reducing their integration density and also introducing unwanted propagation losses.

In this study, we demonstrate integrated photonic filters based on adiabatic structures combining broadband responses with compact footprints through an independent design of band edges in a cascaded structure. We first present a design methodology for compact, single-cutoff, 1×2-port adiabatic filters by optimizing the adiabaticity parameter through a slowly-varying, coupled waveguide structure. The resulting device acts as a single-cutoff filter separating the input signal into short-pass and long-pass outputs. We show that this filter's performance metrics can be controlled by an upper bound on the power coupled out of the fundamental mode, specifically by varying the adiabaticity parameter. The cutoff wavelength of these single-cutoff filters can be adjusted by modifying the cross-sectional geometry of the coupled waveguides. This flexibility allows us to design band-pass filters by cascading two of these single-cutoff filters, designed independently at different cutoff wavelengths. With this configuration, we experimentally demonstrate band-pass filters with bandwidths ranging from 6.4 nm to as wide as 96.6 nm, with inherently flat-top and low-loss transmission.

## II. Device Structure and Design Methodology

The adiabatic filter design uses a coupled waveguide system with two specifically engineered waveguides that are designed to be phase-matched only at a single cutoff wavelength ($\lambda_c$), and phase-mismatched otherwise. The spectral filtering operation is based on placing these two waveguides in close proximity, where the combined structure results in spatial separation of the fundamental modes at short ($\lambda < \lambda_c$) and long ($\lambda > \lambda_c$) wavelengths, on either side of the band edge at the cutoff wavelength. This is made possible through a difference in the group indices of the two waveguides within the desired operation band of the device. For this purpose, our first waveguide is designed using laterally-arranged, multi-segment subwavelength structures with specific effective index and dispersion properties. As demonstrated in Fig. 1a, with increasing number of segments, this first waveguide more closely resembles a uniform material, with a flatter effective index profile. Our second waveguide is a typical strip waveguide, with a significantly different effective index profile. As plotted Fig. 1b, a multi-segment waveguide with more segments has a smaller group index. As such, with more segments, the group index difference between the single-segment and the multi-segment waveguides grows $\Delta n_g^{1-7} > \Delta n_g^{1-5} > \Delta n_g^{1-3}$. This higher difference in the group indices results in better confinement of the fundamental mode to the intended waveguide and enhances filter roll-off. However, with increased number of segments, the two waveguides are separated farther, reducing coupling strength between them, and resulting in longer devices overall. Therefore, we designed our filters using the single-segment ($WG_1$) & five-segment ($WG_5$) waveguide configuration.

Fig. 1c shows the fundamental electric field profiles of this coupled waveguide structure ($WG_1$ & $WG_5$) for wavelengths below and above the filter cutoff. In general, majority of the fundamental mode is spatially confined around the waveguide with the higher effective index. Therefore, the fundamental mode is confined mainly around the single-segment waveguide at short wavelengths and around the multi-segment waveguide at long wavelengths. Such mode confinement onto separate waveguides at different wavelengths enables spectral splitting of short and long wavelengths to respective output ports.

The adiabatic filter is created from these two waveguides ($WG_1$ and $WG_5$), using slowly varying transitions as shown in Fig. 1d. The geometry of the filter is separated into four different adiabatic transitions along the propagation direction, with lengths of $L_1$, $L_2$, $L_3$, and $L_4$. In section 1, $WG_5$ is created by transforming a single-segment input waveguide to the five-segment design. Towards to the end of this section, $WG_1$ is introduced adjacent to $WG_5$. In section 2, the width of the $WG_1$, ($w(z)$) is slowly tapered up. In section 3, $WG_5$ and $WG_1$ are gradually separated, using the separation $s(z)$. Finally, in section 4, $WG_5$ is transformed back into a single-segment output waveguide; and $WG_1$ is also tapered to the correct output waveguide width.

For this design, we used 35 µm-long linear tapers for sections 1 and 4, as these only include transitions within $WG_5$, and near-lossless transmission is easily achieved through typical linear transitions. Outermost adjacent segments are slowly introduced from the sides to limit loss to unwanted modes during the transition (see Appendix A). In contrast, the design of sections 2 and 3 present significant challenges and tradeoffs between device performance and footprint. Specifically, our design variables $w(z)$ and $s(z)$ need to be carefully optimized in order to minimize power lost to the unwanted waveguide, while still maintaining reasonable section lengths. We describe this as an optimization problem in coupled mode theory [48] for adiabatically transforming the fundamental mode while minimizing excitation of the higher order modes.

We use the variable $\gamma = \delta/\kappa$ to express the amount of power in either one of the two waveguides as a function of wavelength, where $\delta = (\beta_5 - \beta_1)/2$ is the mismatch of propagation constants between the individual (uncoupled) modes of $WG_5$ and $WG_1$, and $\kappa$ is the coupling coefficient between these two modes. We model sections 2 and 3 individually, as cascades of successive transitions determined by $w(z)$ and $s(z)$, respectively. As such, the evolution of these parameters along the propagation direction directly influences how the variable $\gamma$ changes throughout the device. This is critical in limiting the unwanted power loss to higher order modes to a minimum, since the adiabaticity criterion indicates that

$$\frac{1}{2\kappa\left(1+\gamma^2\right)^{3/2}}\frac{d\gamma}{dz} \leq \sqrt{\varepsilon} \qquad (1)$$

where $\varepsilon$ is the fraction of power lost from the fundamental mode [35]. Here, we consider power coupling between only the fundamental even and odd modes. By design, as the optical input to our filter geometry is always in the fundamental even mode, the amount of power lost to the odd mode determines how much power is coupled to the adjacent waveguide. The parameter $\varepsilon$ is an upper bound on this power, from which $\gamma(z)$ can be optimized for the shortest possible adiabatic transition. In sections 2 and 3 of our filters, the dimensions and separation between waveguides vary significantly making $\kappa$ also a function of z. For this reason, we use a numerical method for the solution of $\gamma(z)$ in these two sections for a given upper bound $\varepsilon$ as detailed below.

We first select the initial ($w_0$ and $s_0$) as well as final target ($w_T$ and $s_T$) dimensions for each section. Here, $w_T$ is determined based on the desired cutoff wavelength; and $s_T$ is chosen to be approximately 1500 nm so that no significant further coupling can occur between the two waveguides after they are separated. We then choose a design wavelength ($\lambda_d$) slightly shorter than the cutoff wavelength. The design wavelength parameter influences performance metrics of the filters and will be discussed in the following section. Using a mode solver at this design wavelength, we solve for and record the fundamental transverse-electric (TE) mode profile of $WG_1$ as a function of its width, $w$. We also record the fundamental TE mode-profile of $WG_5$, for which the cross-section remains constant throughout these two sections. Using these mode profiles, we created lookup tables by calculating the coupling coefficient $\kappa$ using its overlap integral definition [48] and the corresponding $\gamma$, both as functions of $w$ and $s$. Then, starting from the initial width ($w_0$) for section 2, and the initial separation ($s_0$) for section 3, we discretize and solve Eq. 1 in steps of $\Delta z$ along the propagation direction according to the following algorithm, yielding $w(z)$ and $s(z)$.

given $w_0$ (or $s_0$) at $z = 0$, read $\gamma_0$ and $\kappa_0$ from lookup tables
**while** $w_n < w_T$ (or $s_n < s_T$)
   find the next $\gamma$ by $\gamma_{z+\Delta z} = \gamma_z + 2\kappa_z(1+\gamma_z^2)^{3/2}\sqrt{\varepsilon}\Delta z$
   using $\gamma_{z+\Delta z}$ read $w_{z+\Delta z}$ (or $s_{z+\Delta z}$) from lookup tables
   using $w_{z+\Delta z}$ (or $s_{z+\Delta z}$) read $\kappa_{z+\Delta z}$ from lookup tables
   increment $z = z + \Delta z$
**end**

To illustrate the performance of our filters created with this algorithm, we designed an example adiabatic filter with a five-segment waveguide consisting of 250 nm-wide subwavelength segments separated with 150 nm gaps, and a single-segment waveguide with the target width of $w_T = 308$ nm. We also chose $\lambda_d$ to be 16 nm shorter than the cutoff wavelength. For these parameter selections in section 2, the optimized $w(z)$ for three different $\varepsilon$ bounds are plotted in Fig. 2a. Here, $w(z)$ adiabatically evolves starting from the initial width of $w_0 = 230$ nm, below which we find that there is practically no coupling. In this algorithm, the device length is not subject to any direct constraints, but instead is determined by the selection of $\varepsilon$. As expected, imposing tighter bounds on $\varepsilon$ to limit the power coupled to the unwanted odd mode results in longer transitions. We also observe that $w(z)$ evolves rapidly at the beginning of the section where power coupling is minimal but slows down significantly towards the end as it approaches the target $w_T$ with increased coupling. Similarly for section 3, we plot the optimization result of $s(z)$, from an initial separation of $s_0 = 200$ nm to a final target of $s_T = 1500$ nm in Fig. 2b. Similar to section 2, our algorithm returns longer optimal lengths for this section with tighter bounds on $\varepsilon$. The optimized $s(z)$ evolves slowly at the beginning of the section due to stronger coupling, and changes much faster towards the end as the two waveguides are separated farther.

### III. SIMULATED AND EXPERIMENTAL RESULTS

In order to model the filtering performance of this single-stage filter, we used the transfer matrix method to simulate the adiabatic transitions in sections 2 and 3. Fig. 3a shows the fraction of powers in the fundamental even and odd modes for three separate filters designed with different $\varepsilon$ bounds. As seen in this figure, the fraction of the power in the odd mode is suppressed in the entire spectrum of interest except for the region around the cutoff. This is the spectral region where the difference between the propagation constants of the individual waveguides ($\beta_5$ and $\beta_1$) is the smallest, resulting in increased coupling between these waveguides. Therefore, suppressing the power lost to the odd mode is significantly more challenging around the cutoff, and requires slower transitions. With tighter $\varepsilon$ bounds, our design algorithm creates slower transitions, resulting in a faster decrease of power in the odd mode around the cutoff region, which is necessary for enabling sharp filter roll-offs.

The transmission at the two outputs (short-pass and long-pass) for three different $\varepsilon$ bounds is plotted in Fig. 3b. These results verify the expected flat-top pass-bands due to the inherent near-lossless transmission through the adiabatic structures. Since $\varepsilon$ represents the amount of power allowed to couple to the odd mode, it directly affects the extinction ratio of the filter, where a smaller $\varepsilon$ results in better extinction. The filters also exhibit sharper roll-offs with tighter $\varepsilon$ bounds, as the power in odd mode around the cutoff wavelength shrinks with decreasing $\varepsilon$. We plot this roll-off and the resulting device length ($L_2+L_3$) as a function of the $\varepsilon$ parameter in Fig. 3c. The roll-off here is calculated as the numerical gradient of the transmission at -3 dB. This result clearly illustrates the design tradeoff between filter roll-off and device length. Filters with sharper roll-offs can be obtained by running the algorithm with smaller $\varepsilon$, but at the expense of longer device lengths. The specific selection of the design wavelength ($\lambda_d$) also influences this filter roll-off. As this design wavelength approaches the cutoff wavelength of the filter, $\gamma(z)$ (and the resulting $w(z)$ and $s(z)$) are optimized in order to provide a sharper spectral change in the filter transmission. We observed that as $\lambda_d$ approaches $\lambda_c$, the transition length slightly decreases for section 2, but significantly increases for section 3 (see Appendix B). Once again, this choice also presents a tradeoff between sharper roll-offs and overall device lengths. Considering these tradeoffs between ε or $\lambda_d$ parameters and the resulting device length, our design algorithm can be used to produce filters with varying extinction and roll-off metrics, with lengths ranging from hundreds of micrometers to several millimeters.

One of the most significant advantages of these single-stage adiabatic filters is that they can be individually designed and cascaded, in order to achieve arbitrary filtering profiles in any one of long-pass, short-pass, or band-pass configurations. We demonstrate this by cascading two of these single-stage filters as shown in Fig. 4a to obtain a double-stage, band-pass filter. Here, the individual cutoff wavelengths are denoted by $\lambda_{c1}$ and $\lambda_{c2}$, where $\lambda_{c1}<\lambda_{c2}$. The long-pass output from the first filter is connected to the input of the second filter. Therefore, the short-pass output from the second filter exhibits a band-pass response. We designed each one of these filters by modifying the dimensions of the individual waveguides (specifically $w_T$), in order to place the band edges at the desired cutoff wavelengths. Considering the previously mentioned design tradeoffs, we chose ε ≈ 0.0015 and $\lambda_d$ to be 10 nm shorter than the cutoff wavelength to obtain the best possible extinction and roll-off, while limiting total device length of each individual filter to around 1 mm. Specifically, we designed four double-stage filters with target bandwidths of 5 nm, 20 nm, 50 nm, and 100 nm, each one consisting of two cascaded single-stage filters. We then ran the above algorithm for sections 2 and 3 of each one of these eight single-stage filters. The resulting total device lengths of these independently designed single-stage filters are between 766.8 µm and 1126.4 µm.

The cascaded structures were fabricated using standard 193 nm CMOS photolithography techniques on 220 nm SOI wafers through imec's multi-project-wafer foundry service. An optical microscopy image from the two ends of one of the fabricated band-pass filters is shown in Fig. 4b. Edge couplers (for devices with 50 nm and 100 nm target bandwidths) and grating couplers (for devices with 5 nm and 20 nm target bandwidths) provided by the foundry were used for coupling light into and out of the devices. The fabricated filters were then characterized using a tunable laser source and an optical power meter. A manual 3-paddle polarization controller was used to couple TE light onto the chip.

In Fig. 4c, we plot the experimentally measured transmissions at the three output ports of the double-stage filter with 100 nm target bandwidth. For this filter, we cascaded two single-stage filters with target cutoff wavelengths of $\lambda_{c1}$=1525 nm and $\lambda_{c2}$=1625 nm. Resulting cutoff wavelengths were measured to be $\lambda_{c1}$=1534.6 nm and $\lambda_{c2}$=1631.2 nm. Therefore, this specific filter demonstrates a bandwidth of 96.6 nm at the band-pass output. As predicted by the individual filter simulations of the independent stages, all outputs are spectrally flat across the whole measured spectrum with insertion loss of less than 1.5 dB in all three outputs. The roll-off speeds were measured to be 1.40 dB/nm and 1.55 dB/nm at the two cutoff wavelengths respectively, matching our expectations from the transfer matrix simulations. The extinction ratios were measured to be 14.5 dB, 15.2 dB, and 11.0 dB for the short-pass, band-pass, and long-pass outputs, respectively.

While these extinction ratios are less than those predicted by transfer matrix simulations in Fig. 3b, the difference can be attributed to excitation of higher order modes that are present in our waveguide cross-sections and fabrication imperfections. In theory, what ultimately determines the resulting extinction ratio are two key parameters: γ as a function of wavelength at the end of the device, and how much power (ε) is lost from the fundamental mode by the end of the device. Of these parameters, γ is determined by final waveguide cross-section itself, and ε is specified directly as an input to the optimization algorithm. Therefore, the gradual evolution of waveguide geometry has no direct influence on the extinction ratio, given that this evolution is formed in a way that power lost from the fundamental mode is below a sufficiently small ε and the final cross-section of the structure achieves a sufficiently large γ. However, in our adiabatic design, we limit the analysis to coupling only between fundamental even and odd modes. Power lost to higher order modes may contribute to the power recorded at the other outputs. A more general optimization approach that accounts for coupling loss to all other guided and radiative modes [36] can be used for further improvement of the extinction ratios, depending on specific application requirements. As a generalized version of our two-mode-approach here, look-up tables of coupling coefficients with overlap integrals of all pairs of possible modes can be created. Using this data, the adiabatic transitions can be numerically optimized by limiting the amount of power lost to all of the other modes included in this analysis. By limiting the total optical power coupled out of the fundamental mode, such an approach may result in devices with better extinction ratios than those considering only two modes. Besides the higher order mode coupling, fabrication variations may also contribute to the mismatch of extinction ratios between simulations and experimental measurements. We investigated the effect of potential over-etch under-etch scenarios, and observed deterioration in extinction ratios of about 5 dB, with ±15nm over/under-etch of waveguide widths. In addition to extinction ratio, we also find that the cutoff wavelength deviations of about 10-12 nm from simulation results can be attributed to uneven changes in the widths of $WG_1$ and $WG_5$, again by about 15nm over/under-etch of waveguide widths. These fabrication tolerances and the accompanying analysis can be found in Appendix C.

Measured transmissions of the filters with target bandwidths of 50 nm, 20 nm, and 5 nm are shown in Fig. 4d, 4e, and 4f, respectively. Resulting 3 dB-bandwidths for each one of the filters were 45.5 nm, 25.7 nm, 6.4 nm, respectively. The extinction ratios were again measured to be at least 14.0 dB in the passband; and the insertion losses were measured to be less than 1.5 dB. Except for the narrowest band filter in Fig 4f, all outputs were observed to be spectrally flat. For the device with the bandwidth of 6.4 nm, since the cutoffs of individual filters are spectrally close, resulting double-stage filter can exhibit higher insertion loss due to the filter roll-off (see Appendix D).

Depending on the application requirements, this can be mitigated by improving filter roll-off through optimal ε and $\lambda_d$ selections as necessary, at the expense of longer device lengths. These results demonstrate the flexibility of our design algorithm in creating spectral filters with bandwidths from several nanometers to about 100 nm and beyond, enabling their use in a wide variety of applications.

These band-pass adiabatic filters present an important class of devices enabling FSR-free, ultra-wide passbands with a flexible design procedure, while still providing access to all three outputs in transmission mode for any downstream devices. Crucially, this eliminates the need for the use of a circulator that is required for accessing outputs from devices in reflection mode [29, 30]. Moreover, this new class of filters come with a flexible and scalable design methodology that can be adapted for a variety of applications that require broadband and flat-top transmission responses. In Table I, we compare integrated optical filters of different classes with various performance metrics. Ring resonators are the most compact filters among them, but the widest achievable bandwidths are on the order of several nanometers, even with high-order or cascaded ring schemes due to limited FSRs and underlying resonant dynamics [27]. In the interference-based AWG and MZI lattice filters [20, 25], FSRs determined by the interference order impose an upper bound on the practically usable filter bandwidth, typically up to around 20 nm. Therefore, even though these filters achieve relatively higher extinction, they are significantly limited in terms of bandwidth. These AWG or MZI-based filters typically also require large device footprints on-chip, reducing integration density. With the use of SWGs, the passband bandwidths in grating-based MZI filters have been extended to about 40 nm by increasing the coupling coefficient [28]. However, fabrication of such SWGs can be significantly challenging due to small feature sizes, especially when trying to increase the coupling coefficient beyond what is possible with standard SOI processing capabilities [49]. Prior to our devices here, the widest demonstrated bandwidths have been achieved with contra-directional Bragg couplers, but with much longer device sizes and limited extinction ratios [34]. In these gratings, while a slower chirp rate can improve the extinction in principle, it also greatly increases the device length. In contrast, our waveguide structure does not rely on the coupling between backward and forward propagating modes. This fundamental difference in the underlying filtering approach enables our designs to achieve much wider filter bandwidths than those limited by the strength of grating perturbations. The waveguide geometry we present requires no additional material deposition or fabrication stages that are otherwise used to further increase the coupling coefficient [50]. Moreover, it is also possible to create passbands wider than 100 nm or even more complicated spectral filter shapes using the design procedure we outlined, through flexible and independent design and cascading of the individual stages.

Our filters can be especially useful in applications that require transmission responses wider than typical 100 GHz - 1 THz passbands (0.8nm - 8nm around 1550nm), for which ring resonators or MZIs may be typically sufficient. For instance, ultra-broadband integrated optical (de)multiplexers are essential for scaling the transmission capacity of fiber optic networks. So far, AWGs are commonly used for designing high-density multiplexers utilizing a broad spectrum around the C-band [51-54]. Pinguet et al. have demonstrated wavelength division multiplexer transmission with four channels and with a crosstalk level of -17 dB [55]. Interleaved AWGs [53] and MZI-assisted AWG [54] are also demonstrated as multiplexers with -15 dB of crosstalk in operation, but they suffer from high insertion losses of up to 10 dB. Therefore, owing to its low-loss, flat-top, and broadband operation with superior performance metrics, our filters with double-stage or multi-stage configurations can be utilized in integrated ultra-wideband transmission systems covering the S-, C-, and L- bands. Similarly, the operation bandwidth of spectrometers can be significantly increased by combining multiple bands with our filters. Moreover, in dual-comb spectroscopy applications, spectrally filtering a broadband light source is routinely used to improve the signal-to-noise ratio [56]. Currently, many systems use free-space optical filters to separate the two halves of the dual-comb spectrum [15, 57]. In these applications, our filters can be a scalable on-chip alternative to bulky free-space components, helping reduce system complexity and improve integration density.

## IV. CONCLUSION

In summary, we presented the design of compact integrated spectral filters with ultra-broadband response and low insertion loss through a flexible adiabatic optimization algorithm. We showed that the proposed filters' roll-off and extinction performance can be optimized by modifying the optimization bound on the power lost to unwanted mode in separately designed filter stages. The ability to independently configure band edges through our design algorithm provides the previously elusive design flexibility for filters with different bandwidths. This flexibility also enables other multi-stage filter configurations for implementing more complicated spectral features. Experimentally, we demonstrated band-pass filters composed of two cascaded single-stage filters. Our band-pass filters experimentally achieve bandwidths up to 96.6 nm with flat-top transmissions, a maximum insertion loss of 1.5 dB, and an extinction ratio of over 15 dB in the passband. With the widest bandwidths reported to date, these devices can enable new possibilities in next-generation, high-throughput, and ultra-broadband communications and spectroscopy applications.

## APPENDIX

### A. FDTD analysis of the transitions in sections 1 and 4

In sections 1 and 4 of the adiabatic filter design, the single-segment waveguide and the five-segment waveguide are slowly evolved into one another through adiabatic transitions. One possible design for these transitions involves using linear tapers from the single-segment waveguide to the five-segment waveguide as shown in Fig. 5a. However, abrupt introduction of new adjacent segments can potentially lead to undesired losses. Using the 3D finite difference time domain (FDTD) method we calculated the transmitted power of the fundamental mode for this transition. We used fundamental TE mode as the input and then calculated the amount of power transmitted to the fundamental mode at P-plane. As shown in Fig. 5c (black line), more than 1% of the power is lost within this transition. In order to reduce this loss, we utilized adiabatic bends when introducing the outermost segments in this transition as shown in Fig. 5b. Transmission plot (red line in Fig. 5c) shows that the amount of power lost is significantly reduced when the adjacent waveguides are introduced to the structure through these bends.

### B. Effect of design wavelength ($\lambda_d$) on adiabatic filter design

For a given wavelength, which we call the design wavelength ($\lambda_d$), Eq. 1 can be solved and $\gamma(z)$ can be obtained. However, since $\gamma$ changes with wavelength, solution of Eq. 1 at different $\lambda_d$ will result in a different $\gamma(z)$. In order to analyze how the choice of $\lambda_d$ affects the adiabatic transitions, we fixed the $\varepsilon$ bound at 0.005, and calculated $w(z)$ and $s(z)$ for different $\lambda_d$. We used a five-segment waveguide consisting of 250 nm wide subwavelength layers separated with 150 nm gaps. We chose the target width of single-segment waveguide as 308 nm resulting in a cutoff wavelength ($\lambda_c$) of 1546 nm. For convenience, we defined a new parameter, $\Delta\lambda = \lambda_c - \lambda_d$, as the difference between the cutoff wavelength and the design wavelength. In Fig. 6a and Fig. 6b, we plot the calculated $w(z)$ (for section 2) and $s(z)$ (for section 3), for three different choices of $\Delta\lambda$ ranging from 6 nm to 26 nm. From these results, we conclude that as $\Delta\lambda$ decreases (i.e., $\lambda_d$ approaches $\lambda_c$), transition length decreases for the width, $w(z)$, but significantly increases for the separation, $s(z)$.

To better understand this behavior, we rearranged Eq. 1 as

$$\frac{d\gamma}{dz} \leq 2\kappa(1 + (\delta/\kappa)^2)^{\frac{3}{2}}\sqrt{\varepsilon} \qquad (2)$$

and plotted the right-hand-side (RHS) of Eq. 2 in Fig. 6c as a function of WG$_1$'s width $w(z)$ in section 2. This RHS serves as an upper bound for $d\gamma/dz$, and is shown again for the three different choices of $\Delta\lambda$. From this analysis, we note that as the width of WG$_1$ increases, $max$ ($d\gamma/dz$) decreases to a minimum and then increases again. That is because with a wider WG$_1$, the phase mismatch between the waveguides decreases at the beginning and becomes zero at the width where two waveguides are phase-matched at corresponding $\Delta\lambda$. Here, the width for WG$_1$ at which the two waveguides are phase-matched is a function of the design wavelength, and therefore is also a function of $\Delta\lambda$. When WG$_1$'s width is greater than the phase matching width, there is greater phase mismatch; and thus $max$ ($d\gamma/dz$) increases again. Therefore, the transition becomes much slower as the width of WG$_1$ approaches the point where two waveguides are phase-matched. Comparing different $\Delta\lambda$, we find that $max$ ($d\gamma/dz$) is greater for smaller $\Delta\lambda$ at a majority of the width range up to 308 nm (260 nm - 305 nm), which enables shorter transitions for smaller $\Delta\lambda$. This can be explained again by phase-mismatch between the two waveguides. For the widths between 260 nm and 305 nm, with longer design wavelengths (i.e. smaller $\Delta\lambda$) the phase mismatch between the waveguides grows, resulting in a higher $max$ ($d\gamma/dz$). For example, at a width of 300 nm, the phase-matching wavelength is around 1500 nm. Therefore, at this width, phase mismatch between waveguides is the greatest for $\Delta\lambda$ = 6 nm ($\lambda_d$ = 1540 nm) resulting in the highest $max$ ($d\gamma/dz$), and is the smallest for $\Delta\lambda$ = 26 nm ($\lambda_d$ = 1520 nm) resulting in the lowest $max$ ($d\gamma/dz$).

For the same three $\Delta\lambda$, we plot the maximum bound on $d\gamma/dz$ in Fig. 6d, this time as for section 3 as a function of the separation $s(z)$ between WG$_1$ and WG$_5$. Here, the coupling coefficient decreases with increasing separation; and therefore $\kappa \to 0 \Rightarrow 2\kappa(1 + (\delta/\kappa)^2)^{3/2} \to 2\,\delta^3/\kappa^2$. Hence, $max$ ($d\gamma/dz$) increases with decreasing κ, making the transitions faster as separation increases. Comparing different $\Delta\lambda$, we find that as $\Delta\lambda$ decreases, $max$ ($d\gamma/dz$) is significantly suppressed due to the two waveguides becoming more closely phase-matched. This drop in $max$ ($d\gamma/dz$) results in much longer transitions, in order to maintain the same amount of maximum power ($\varepsilon$) coupled to the adjacent waveguide, at design wavelengths much closer to the cutoff wavelength.

Using the transfer matrix method, we simulated the structures obtained from the results in Fig. 6. First, we plot the fraction of power in the fundamental even and odd modes in Fig. 7a. This power is calculated at the end of the filter designed with the three different $\Delta\lambda$. With smaller $\Delta\lambda$, while the power in the odd mode remains constant at the cutoff wavelength $\lambda_c$ itself, it is significantly reduced on either side the cutoff. This is evident from the faster spectral roll-off of the odd mode power on either side of the cutoff wavelength. This behavior is expected as the adiabatic criterion ensures that the amount of power lost to the odd mode is kept at a minimum at the design wavelength $\lambda_d$. Therefore, choosing a design wavelength closer to the cutoff wavelength suppresses the power in the odd mode around the cutoff.

Fig. 7b shows the transmission spectra of these filters. We observe that as $\Delta\lambda$ decreases, the filter roll-off becomes sharper, at the expense of longer overall device lengths. These sharper roll-offs can be beneficial for spectral separation in many applications. However, we also note that using a smaller $\Delta\lambda$ also decreases the extinction ratio of the filter. This can be explained by the design algorithm's tendency to strongly favor the spectral separation and a sharp filter roll-off when using a smaller $\Delta\lambda$ choice. This problem can be prevented by a more general adiabatic design process where multiple design wavelengths are used, and the device optimization procedure limits higher mode excitation at each one of these wavelengths.

### C. Effect of width deviation on the filter response

In this section, we investigated the effect of fabrication variations on the filter performance. We chose a filter with a cutoff wavelength ($\lambda_c$) of 1546 nm as our reference structure, which has a five-segment waveguide consisting of 250 nm wide subwavelength segments separated by 150 nm gaps, and a single-segment waveguide with 308 nm target width. We calculated the optimized adiabatic transitions of section 2 -$w(z)$- and 3 -$s(z)$- by setting $\lambda_d$ as 1530 nm and $\varepsilon$ as 0.005. Then we performed the transfer matrix simulations of the same structure under various over-etch and under-etch amounts. Here, we consistently increased (decreased) the width of each subwavelength segment of the five-segment waveguide and width of the single-segment waveguide by certain under-etch (over-etch) scenarios, indicated below by the variable $\Delta w$. The center-to-center separation of the waveguide segments remained unchanged in this analysis. Fig. 8 shows the simulated transmission spectra at long-pass outputs in (a) under-etch ($\Delta w > 0$) and (b) over-etch ($\Delta w < 0$) cases. In this case, since both five-segment and single-segment waveguides are subject to the same amount of etch offsets, the phase matching wavelength of $WG_1$ and $WG_5$ only changes slightly. For instance, for 10 nm of under-etch and over-etch cases, cutoff wavelength shifts from the ideal device are observed to be less than 1 nm and around 2 nm, respectively. A blue-shift of cutoff is observed in both under-etch and over-etch scenarios, which can be explained by the effective index of $WG_5$ being greater than that of $WG_1$ both when $\Delta w > 0$ and $\Delta w < 0$, resulting in phase matching of the two waveguides at a shorter wavelength. In addition to the slight change in the cutoff wavelength, we also observe that these width variations worsen the extinction ratio by about 5 dB. This degradation is a direct result of the violation of the adiabaticity condition in Eq. 1 due to the modified profile of $\gamma(z)$, resulting from the changes in the propagation constants of $WG_1$ and $WG_5$ when $\Delta w \neq 0$.

In our experimental measurements, we observed a redshift of cutoff wavelength by about 10 nm. In order to model this, we investigated several scenarios where $WG_1$ and $WG_5$ are subject to different etch offsets (specifically where $\Delta w_{WG_1} > \Delta w_{WG_5}$). Such uneven etch offsets of the two waveguides more strongly influences the phase matching wavelengths of the two waveguides. As shown in Fig. 9, an increasing difference between $\Delta w_{WG_1}$ and $\Delta w_{WG_5}$ results in a worse measured extinction ratio and emerging sidelobes in the transmission spectrum, similar to the spectral features we measured experimentally in our double-stage filters in Fig 4.

## D. Effect of roll-off on bandwidth

The flexibility of designing individual spectral filters allows building more complicated transfer functions using multiple different single-stage devices in combination. For bandpass devices, two of these single-stage filters with appropriate cutoff wavelengths can be cascaded in series. Fig. 10 shows the passband configurations of cascaded filters for two different cases, depending on the spectral separation of the cutoff wavelengths of single stage filters ($\lambda_{c1}$ and $\lambda_{c2}$). In Fig. 10a, we indicate that if the cutoff wavelengths of the two single stage filters are sufficiently different, then a flat and inherently lossless passband can be obtained in the bandpass output. Bandwidth of this cascaded filter is then simply calculated as the difference between the cutoffs, namely $\lambda_{c2} - \lambda_{c1}$.

However, if the cutoff wavelengths of individual filters are chosen to be spectrally close as shown in Fig. 10b, the bandpass output of the resulting double-stage filter experiences some insertion loss. We denote the individual filter roll-offs with $\alpha$ (dB/nm) and their spectral intersection with a transmission of $-\chi$ (dB). The design procedure we outlined provides flexibility in the roll-off speed $\alpha$ through the choice of the design wavelength with respect to the cutoff wavelength. Given $\alpha$, then the transmission $-\chi$ at the spectral intersection of the two filters is simply a function of how their cutoff wavelengths are positioned spectrally. With these parameters, the overall device exhibits an insertion loss of $2\chi$ dB, and a 3dB-bandwidth of $(2\chi+6)/\alpha$ nm, respectively. It is possible to eliminate insertion loss completely, by careful selection of the cutoff wavelengths to be sufficiently far from one another. In that case, bandwidth is only limited by the roll-off speed (if $\chi=0$, then BW=$6/\alpha$ nm). Therefore, in order to achieve narrowband filters, single-stage filters with sharper roll-offs (higher $\alpha$) are required. As an example, if the roll-off speed is 1dB/nm for both single stage filters, then the narrowest achievable 3 dB-bandwidth for the band-pass filter is 6 nm. This estimate can be used as a starting point for filter design using the approach we outline. For a more detailed analysis of the filter transfer function and filter bandwidth, the transfer matrix method described in the manuscript yields more accurate results since the filter roll-offs are not perfectly linear in the spectral domain, and the two filters can exhibit slightly different roll-offs than each other.

ACKNOWLEDGMENT

E. S. M. acknowledges support by The Science Academy (BAGEP), Turkey. K. G. acknowledges TUBITAK-BIDEB for PhD scholarship.

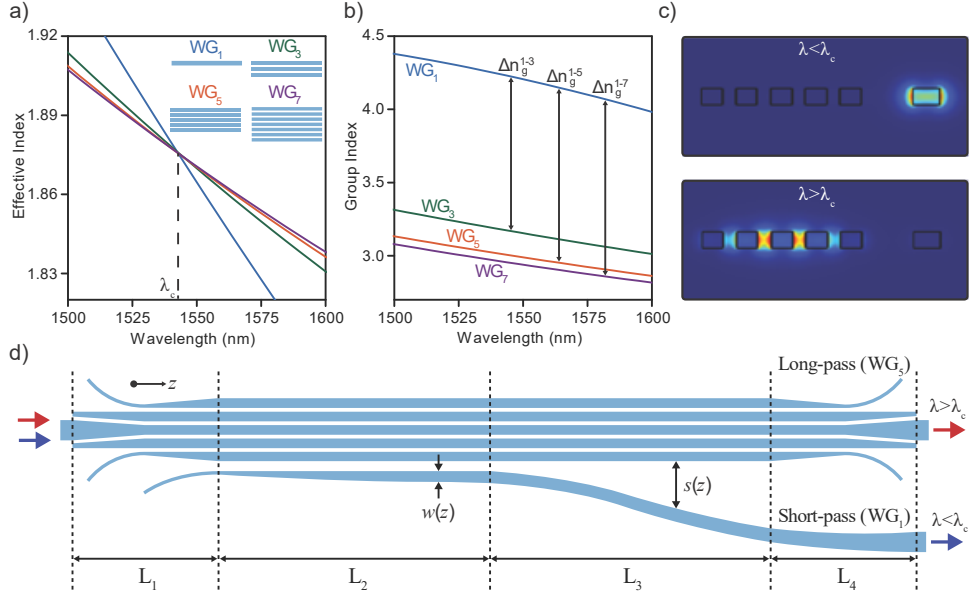

Fig. 1. (a) Effective and (b) group indices of the fundamental transverse-electric mode of the single-segment (WG$_1$), three-segment (WG$_3$), five-segment (WG$_5$), and seven-segment (WG$_7$) waveguides. The width of the single-segment waveguide is 308 nm, widths of individual subwavelength layers of three-, five-, and seven-segment waveguides are 262 nm, 250 nm, and 247 nm, respectively. Gaps between the subwavelength layers are 150 nm. (c) Fundamental transverse electric mode profiles of WG$_1$ & WG$_5$ coupled waveguide configuration at wavelengths below and above the cutoff wavelength ($\lambda_c$). (d) Schematic of an adiabatic single-stage filter composed of coupled single-segment and five-segment waveguides. L$_1$, L$_2$, L$_3$, and L$_4$ are the lengths of four different transitions along the propagation direction. $w(z)$ and $s(z)$ are the width and separation of sections 2 and 3, respectively.

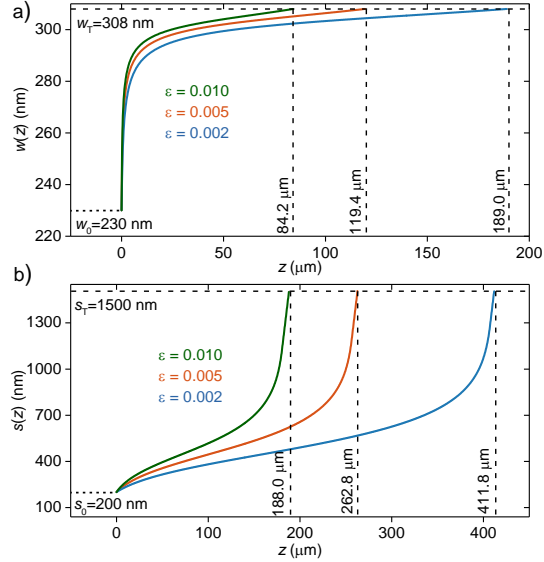

Fig. 2. Optimized (a) width $w(z)$ for section 2 and (b) separation $s(z)$ for section 3, using three different ε bounds. The coupled waveguide system consists of a five-segment waveguide with 250 nm wide subwavelength segments separated by 150 nm gaps, and a single-segment waveguide with the target width of 308 nm.

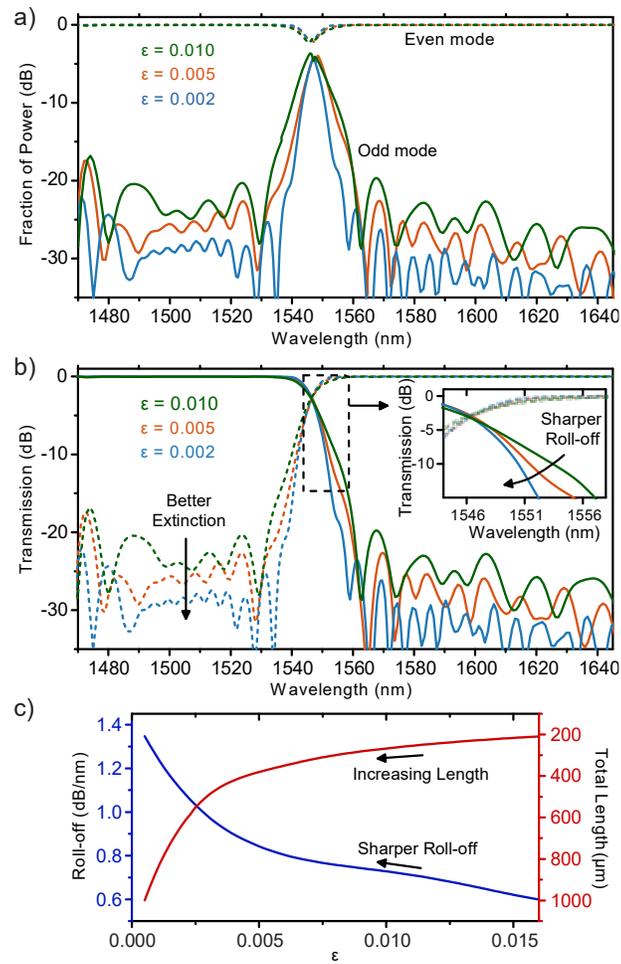

Fig. 3. (a) Simulated fraction of power in fundamental even and odd modes of the optimized single-stage filter designed with three different ε bounds. (b) Simulated transmission spectra of the single-stage filter with three different ε bounds. Inset shows close-up of the filter roll-off. Solid and dashed lines represent short-pass and long-pass outputs, respectively. (c) Roll-off and Total Length as a function of ε. Total Length is the sum of the adiabatic transition lengths for sections 2 and 3 ($L_2+L_3$).

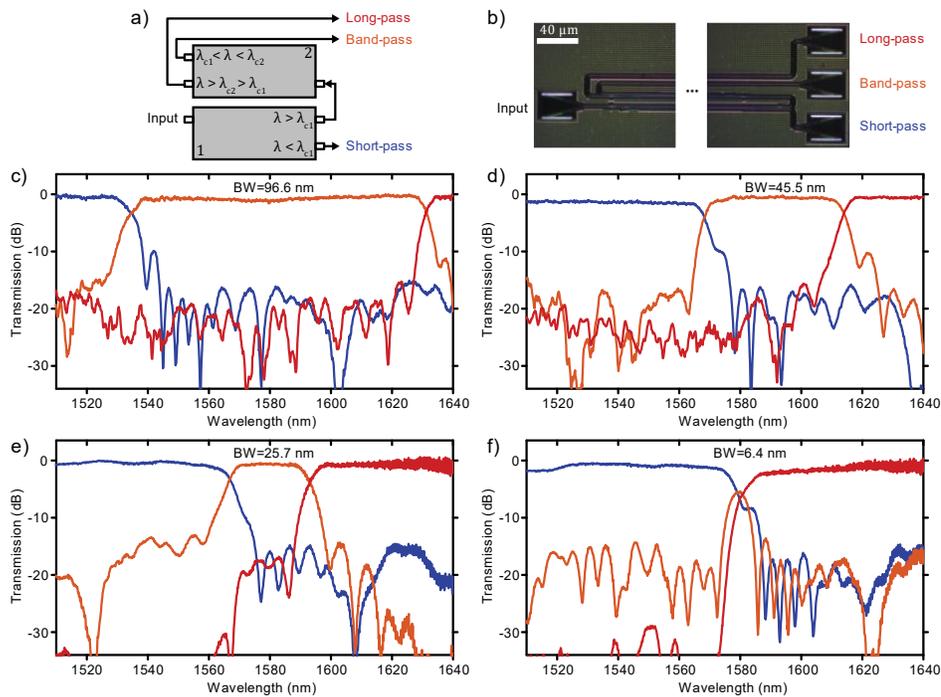

Fig. 4. (a) Block diagram of a double-stage filter composed of two cascaded filters with cutoff wavelengths of $\lambda_{c1}$ and $\lambda_{c2}$. (b) Optical microscopy images taken from the input and output sections of a double-stage filter. Measured transmission spectra of the double-stage filters with bandwidths of (c) 96.6 nm, (d) 45.5 nm, (e) 25.7 nm, (f) 6.4 nm. Blue, orange, and red lines represent short-pass, band-pass, and long-pass outputs, respectively.

TABLE I. Comparison and performance metrics of integrated optical filters.

| Filter Type | Insertion Loss (dB) | Extinction Ratio (dB) | 3dB-Bandwidth (nm) | Bandwidth Limitation | Size or Length |
|---|---|---|---|---|---|
| High-order ring resonator [27] | 1.0 | 25.0 | 2.0 | FSR, resonant nature | 10×30 μm² |
| AWG [25] | 2.0 | 19.1 | ~10.0 | FSR | 305×260 μm² |
| MZI lattice [20] | 1.0 | 20.0 | 19.0 | FSR | 300×100 μm² |
| SWG-assisted MZI [28] | 2.0 | 15.0 | ~40.0 | coupling coefficient | 6 mm |
| Chirped contra-DC [34] | 1.77 | 9.7 | 88.1 | coupling coefficient | 4.7 mm |
| This work (cascaded filter) | 1.5 | 15.2 | 96.6 | no fundamental limitations | 8.6×1083.8 μm² |

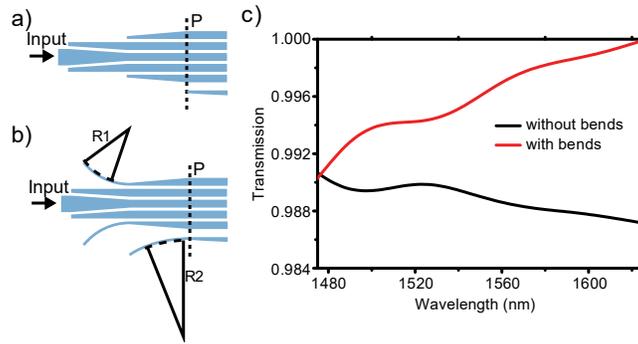

Fig. 5. Schematic representation of single-segment to five-segment transition without (a) and with (b) adiabatic bends. Radii of curvature are: R1=40 µm and R2=80 µm. (c) 3D FDTD-simulated transmission results of the transitions in (a) and (b).

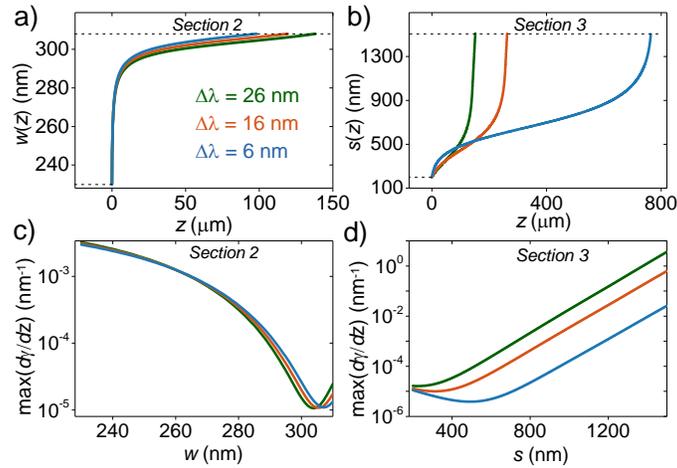

Fig. 6. Calculated (a) $w(z)$ and (b) $s(z)$ for three different $\Delta\lambda$. Calculated RHS of Eq. 2 as a function of (c) $w$ and (d) $s$. ($\Delta\lambda = \lambda_c - \lambda_d$)

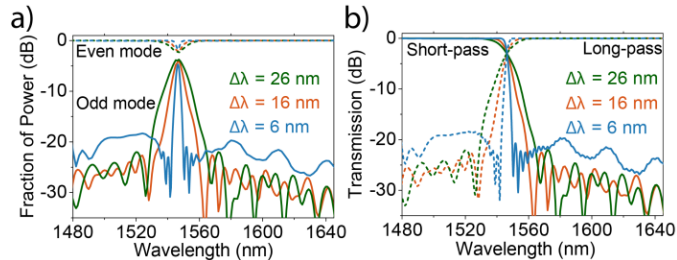

Fig. 7. (a) Simulated fraction of power in fundamental even and odd modes at three different $\Delta\lambda$. (b) Simulated transmission spectra of the single-stage filters designed at three different $\Delta\lambda$. ($\Delta\lambda = \lambda_c - \lambda_d$)

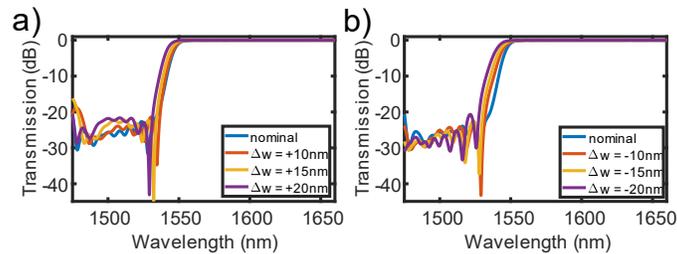

Fig. 8. Simulated long-pass transmission spectra of the filter in (a) under-etch and (b) over-etch cases. Widths of both $WG_1$ and $WG_5$ are modified by the same amount, indicated by $\Delta w$.

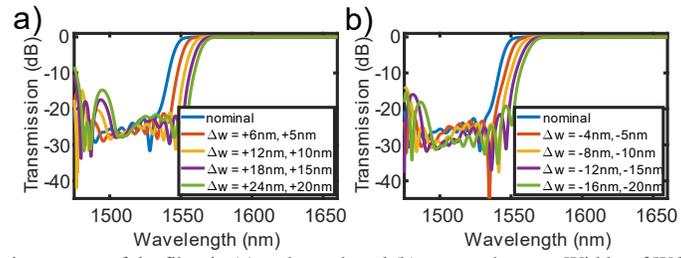

Fig. 9. Simulated long-pass transmission spectra of the filter in (a) under-etch and (b) over-etch cases. Widths of $WG_1$ and $WG_5$ are modified by different amounts. In the legends, the first and second values represent width deviations for $WG_1$ and $WG_5$, respectively.

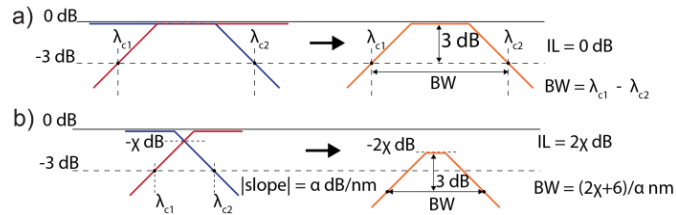

Fig. 10. Passband configurations of cascaded filters for (a) $\lambda_{c1}$ and $\lambda_{c2}$ are spectrally distant, (b) $\lambda_{c1}$ and $\lambda_{c2}$ are spectrally close. (IL: insertion loss, BW: 3dB-bandwidth)